\documentclass{caosp309}

\usepackage{hyperref}
\usepackage{multirow}
\usepackage{arydshln}
\usepackage{rotating, float, caption}
\usepackage{setspace}   
\usepackage[normalem]{ulem}

\usepackage{graphicx}%
\usepackage{multirow}%
\usepackage{amsmath,amssymb,amsfonts}%
\usepackage{amsthm}%
\usepackage{mathrsfs}%
\usepackage{xcolor}%
\usepackage{textcomp}%
\usepackage{manyfoot}%
\usepackage{booktabs}%
\usepackage{algorithm}%
\usepackage{algorithmicx}%
\usepackage{algpseudocode}%
\usepackage{listings}%
\usepackage{url}%
\usepackage{rotating}%
\usepackage{lscape}%
\usepackage{adjustbox}%
\usepackage{tablefootnote}%
\usepackage{booktabs,enumitem}%
\usepackage[footnotehyper]{nicematrix}%
\usepackage{hyperref}%
\usepackage{threeparttable}%

\usepackage{natbib}
\articleNo{301}
\pubyear{2024}
\volume{50}
\volnumber{3}
\firstpage{1}
\received{May 1, 2024}
\accepted{July 28, 2024}

\def\BibTeX{{\rm B\kern-.05em{\sc i\kern-.025em b}\kern-.08em
             T\kern-.1667em\lower.7ex\hbox{E}\kern-.125emX}}

\begin{document}

%
\htitle{Asteroid (12499) 1998 FR47 -- binary or not?}
\hauthor{G.\,Borisov {\it et al.}}

\title{The possible dual nature of the asteroid (12499) 1998 FR47}


%
%
\author{
        G.\,Borisov\inst{1,2}\orcid{0000-0002-4516-459X}
      \and
        N.\,Todorovi\'c\inst{3} 
      \and 
        E.\,Vchkova-Bebekovska\inst{4}   
      \and 
        A.\,Kostov\inst{1}
      \and 
        G.\,Apostolovska\inst{4}
       }

%
\institute{
           Institute of Astronomy with NAO, Bulgarian Academy of Sciences,\\72 Tsarigradsko Chauss\'ee Blvd, BG-1784, Sofia, Bulgaria,\\ \email{gborisov@astro.bas.bg}
         \and 
           Armagh Observatory and Planetarium, College Hill,\\Armagh BT61 9DG, Northern Ireland, United Kingdom 
         \and 
           Astronomical Observatory, Volgina 7, Belgrade, 11060, Serbia 
         \and
           Institute of Physics, Ss. Cyril and Methodius University in Skopje,\\Faculty of Natural Sciences and Mathematics-Skopje, Arhimedova 3,\\Skopje, 1000, Republic of Macedonia
          }

\date{June 26, 2024}

\maketitle


\begin{abstract}
We present the R-band lightcurves of the Flora family asteroid (12499) 1998 FR47, obtained in 2022 at two different astronomical sites: Bulgarian National Astronomical Observatory Rozhen (MPC Code 071) and Astronomical Station Vidojevica (MPC Code C89). The quadramodal lightcurves reveal a rotation period of 6.172$\pm$0.003\,h and an amplitude of about 0.44\,mag. Using the lightcurve inversion method, with the combination of our dense lightcurves and the sparse data from Gaia DR3, we found the sidereal period, an indication of a retrograde rotation of (12499) and its low-convex resolution shape. 
Nonetheless, the unusual shape of the quadramodal lightcurve and its additional analysis reveals two possible periods, 3.0834$\pm$0.0085\,h and 4.1245$\pm$0.0151\,h, making the suspect that the asteroid might be a non-syn\-chro\-nised wide binary system. 
Spectral analysis of the asteroid using data from the GAIA DR3 shows that it is either an M- or an L-type object and maybe a piece of the first planetesimals to form in the Solar System protoplanetary disk. On the other hand, (12499)'s dynamical properties indicate a significantly shorter lifetime. The asteroid lies exactly on the chaotic border of the 7:4 mean motion resonance with Mars (7M:4), alternating between being in and out of it for almost 190 Myrs. During 200\,Myrs of integration, (12499) visited other resonances in the Flora family, but it never became a Near Earth Object (NEO). Additional integration of fictive objects from the 7M:4 resonance showed a possibility of transportation to the NEO region already at about 20\,Myrs. 
\end{abstract}


\section{Introduction}\label{sec:intro}

An asteroid lightcurve is constructed by the rotation of an irregular ellipsoidal shape. In the usual case, lightcurves during one rotational phase should have two maxima and two minima with slightly different depths.
On the other hand, some asteroids have shown
unusual lightcurve shapes, which may be caused by different albedo over their surfaces or irregular shapes. Lately, different observational techniques confirm that some of these irregular lightcurves shapes could be caused by the binary nature of the observed asteroids \citep{Tedesco,Margot2015}. 
In the last two decades, numerous studies of these asteroid binary systems have been done, and the estimation is that around 15\,\% of the Near Earth Asteroids (NEA's), with diameters greater than 0.3\,km, belong to this class \citep{Pravec2006}. There are a few characteristics of the lightcurve that may suggest the possibility of a binary system. The most common one is the presence of two minima with different depths separated by a half phase, which is usually caused by mutual events like occultations and eclipses between the two components. Unfortunately, determining a binary system is not always straightforward. If the system is not positioned equatorially to the observer or the components of the binary are distinctly separated, these features will not be presented in the lightcurve.

The asteroid (12499) 1998 FR47 was discovered on 1998 March 20 by the LINEAR project in Soccoro (New Mexico, USA). According to the Asteroid Lightcurve Photometry Database (ALCDEF)\footnote{\url{https://alcdef.org}} \citep{ALCDEF,ALCDB}, the calculated diameter of (12499) is 4.591\,km, and the assumed albedo is 0.277 \citep{NEOWISE}. A search of the Asteroid Lightcurve Database found no previously reported period for asteroid (12499). The only available information about the asteroid lightcurve was its amplitude of 0.4\,mag, reported by \cite{2013MPBu...40..146S}.  

The asteroid (12499) is located at about $2.21 \,AU$ in the inner main belt in the region of the Flora family (although its membership to the family is somewhat questionable).  According to  \cite{2020PDS...N} it belongs to the Flora family, while the AstDys  database\footnote{Available at \hyperlink{https://newton.spacedys.com/astdys/}{https://newton.spacedys.com/astdys/}} denoted it as a background asteroid. This discrepancy can be explained by the fact that the Flora family does not have sharp boundaries, but it gradually fades into the background population (see e.g. \citet{2017AJ....153..172V, 2014Icar..243..111D}). Nevertheless, here we study only one of its potential members in the same way we studied the asteroid (4940) Polenov \citep{2021SerAJ.202...39V}.

\section{Observations and data reduction}\label{sec:obs}

Our first photometric observations of the asteroid (12499) 1998 FR47 were on 2022 Aug 18 and 19, at Bulgarian National Astronomical Observatory (BNAO) Rozhen (MPC Code 071), when our primary targets were long-term studied asteroids. The observations were performed with a 50/70 Schmidt telescope, equipped with an FLI PL 16803 4096$\times$24096 CCD with 9\,$\mu$m pixels. 
The next observations were conducted on 2022 Sep 23 and 24 at Astronomical Station (AS) Vidojevica (MPC Code C89) with 60\,cm Cassegrain "Nedeljkovi\'c" telescopes equipped with an SBIG STXL-6303 3072$\times$2047 CCD with 9\,$\mu$m pixels.  

After light image reduction, aperture photometry of the asteroid and comparison stars were performed with the software program CCDPHOT\footnote{\url{https://www.boulder.swri.edu/~buie/idl/ccdphot.html}} by \citet{BUIE}. For the lightcurve analysis, we used the software package MPO Canopus v10.8.4.1 \footnote{\url{https://minplanobs.org/BdwPub/php/mpocanopus.php}}, which produces composite lightcurves, calculates rotational periods, provides the Fourier analysis fitting procedure, and estimates the amplitude of the lightcurve.

\begin{figure*}[h]%
\centering
\includegraphics[width=0.99\textwidth]{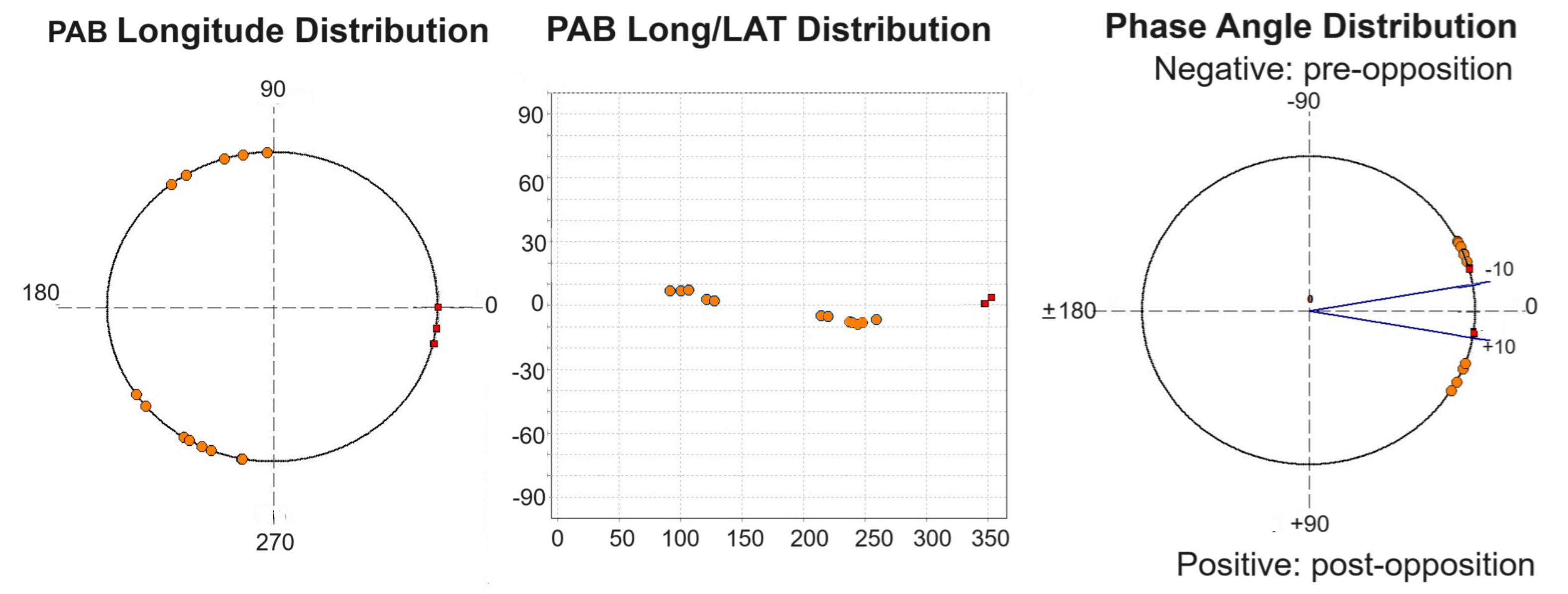}
\caption{Dense (red points) and Gaia DR3 sparse observations (orange points). Left: Distribution of PAB longitude; Centre: Distribution of PAB longitude on PAB latitude; Right: Distribution of phase angle.}\label{fig:aspect}
\end{figure*}

Table~\ref{tab1} presents the aspect data for (12499) for the observational sets and Fig.~\ref{fig:aspect} shows the Phase Angle Bisector (PAB) longitude, PAB latitude, and phase angle distributions for dense and sparse photometric data. Dense data correspond to observations in Table~\ref{tab1} and sparse data correspond to 19 accurate photometric points from 31 December 2014 to 13 May 2017 from GAIA Data Release 3 (DR3)\citep{Gaia,GaiaDR3,GaiaSS}. The phase angles of the sparse satellite data are between 19 and 32 degrees and the phase angles of our dense data fill the gap around the asteroid opposition.

\begin{table}[h]
\centering
\begin{threeparttable}
\caption{
Aspect data for the asteroid (12499) 1998 FR47 for the five observational sets. 
}\label{tab1}
\begin{tabular}{ccccccc}
\toprule%
Date\tnote{1} & $r$\tnote{2} & $\Delta$\tnote{3} & Ph. angle\tnote{4} & $\lambda_{e}$\tnote{5} & $\beta_{e}$\tnote{6} & MPC\tnote{7}\\
(UT) & (AU) &  (AU) &  ($\alpha^\circ$)  &  ($^\circ$) & ($^\circ$) & code\\
\midrule
2022 Aug 19.02  & 1.707 & 0.747 & 16.7016 & 355.04 & 0.26 & 071\\
2022 Aug 20.01  & 1.707 & 0.746 & 16.1411 & 354.96 & 0.40 & 071\\
2022 Sep 23.99 & 1.713 & 0.721 &  7.6643 & 348.70 & 5.12 & C89\\
2022 Sep 24.04 & 1.713 & 0.721 &  7.7522 & 348.67 & 5.14 & C89\\
2022 Sep 24.83 & 1.714 & 0.724 &  8.2522 & 348.53 & 5.22 & C89\\
\bottomrule
\end{tabular}
\begin{tablenotes}
\item[1] The observation date referring to the mid-time of the lightcurve observed
\item[2] Heliocentric distance
\item[3] Geocentric distance
\item[4] The solar phase angle
\item[5] The J2000.0 ecliptic longitude
\item[6] The J2000.0 ecliptic latitude
\item[7] Minor Planet Centre Observatory Code \\ \url{https://www.minorplanetcenter.net/iau/lists/ObsCodesF.html}
\end{tablenotes}
\end{threeparttable}
\end{table}

\section{Results}\label{sec:res}
\subsection{Lightcurve analysis}\label{sec:LC}
During the first night (12499) was observed for about 2.8\,hours. By visual inspection of the typical asteroid bimodal symmetrical lightcurve, we thought that more than 90\,\% of its period was covered. To our surprise, the next night's observations with a duration of 4.5\,h showed an asymmetrical lightcurve having three minima with different amplitudes. Combining these two nights' observations, we could not determine the rotational period.

\begin{figure}[h]%
\centering
\includegraphics[width=\textwidth]{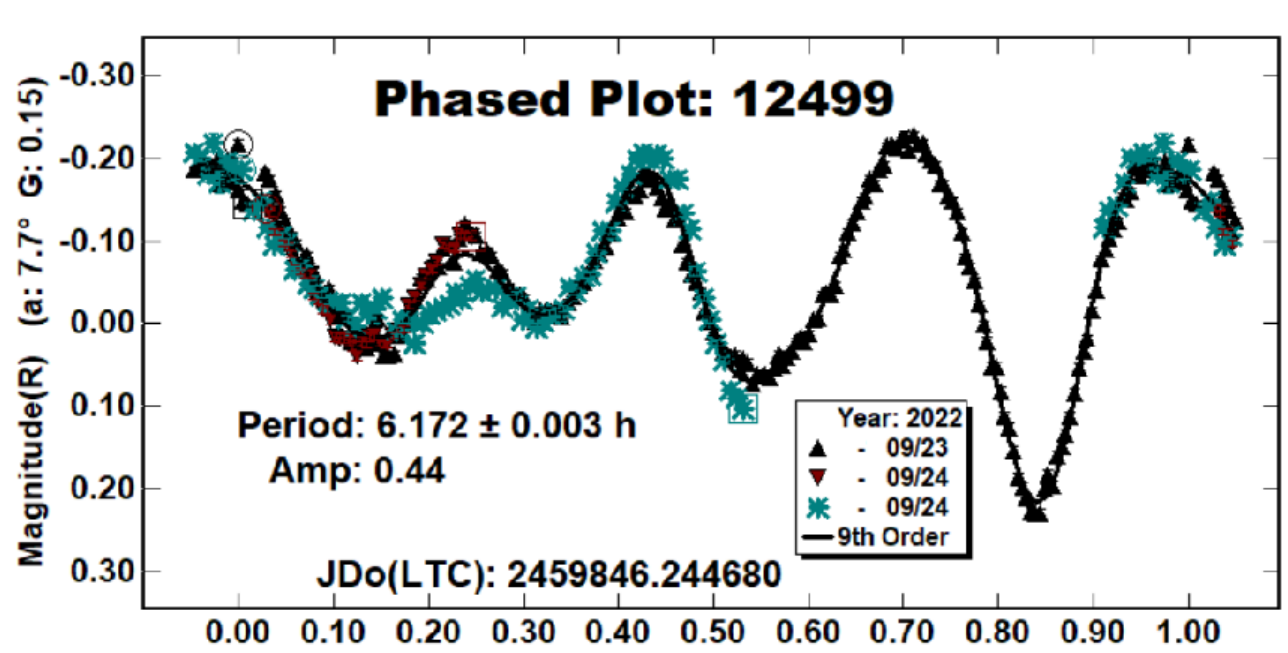}
\caption{The composite lightcurve of (12499) 1998 FR47 gathered from the observations on 23 and 24 September 2020. The Fourier analysis of
order 9 gave the best corresponding to the rotational period 6.172h$\pm$0.003\,h with the amplitude of 0.44$\pm$0.01\,mag.}\label{fig1}
\end{figure}

During the next observations, carried out at AS Vidojevica, Sep 23 and 24, 2022, the solar phase angle of the asteroid was at 7.8$^\circ$ after opposition, and the asteroid was observed for about 8 and 4 hours, respectively. A Fourier fit of order 9 reveals an asymmetric quadramodal shape of the composite lightcurve and confirms a period of 6.172$\pm$0.003\,h with an amplitude of 0.44$\pm$0.01\,mag (see Fig.~\ref{fig1}). 
Using the empirical relation between the amplitude and largest to the smallest axis ratio (a/c) proposed by \citet{Amp0} we computed a/c=1.39. 
Four well-defined maxima and minima with different heights could be due to an irregular shape of the asteroid and/or albedo variations on its surface. 

\begin{figure}[h]%
\centering
\includegraphics[width=\textwidth]{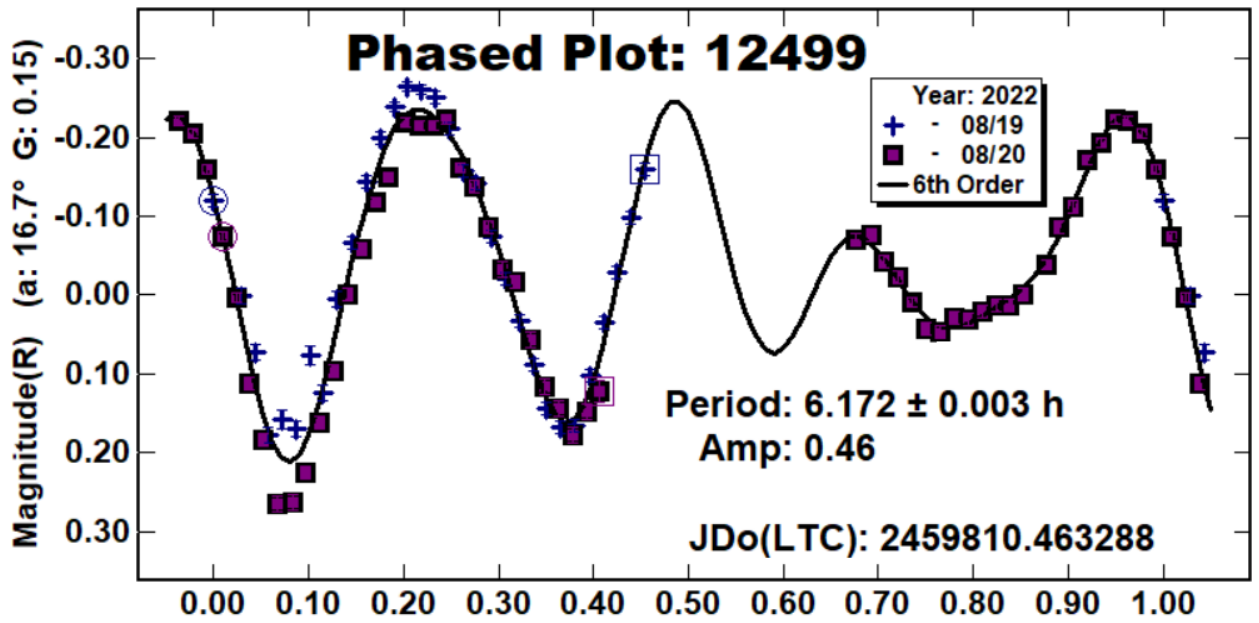}
\caption{The lightcurve of (12499) 1998 FR47 gathered from the observations in August 2022. Although not completely covered, it is constructed on the basis of the afterwards calculated period of 6.172\,h.}\label{fig2}
\end{figure}

This period was used in constructing the composite lightcurve from the previous observations, which were taken at a phase angle of 16.7$^\circ$ before opposition. Even though the data have a gap, the composite lightcurve constructed with a Fourier fit of order 6 again reveals an asymmetric quadramodal shape, that looks like a mirror image of the first composite lightcurve (see Fig.~\ref{fig2}).

\subsection{Testing a single-body shape model}\label{sec:shape}

\begin{figure}[h]%
\centering
\includegraphics[width=0.49\textwidth]{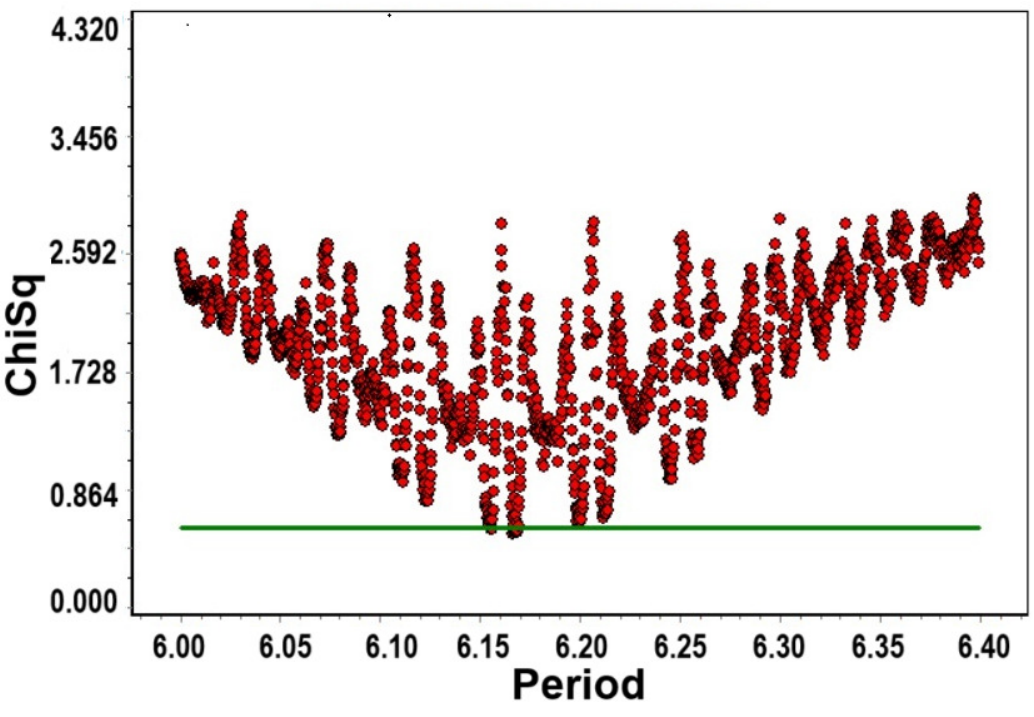}
\includegraphics[width=0.49\textwidth]{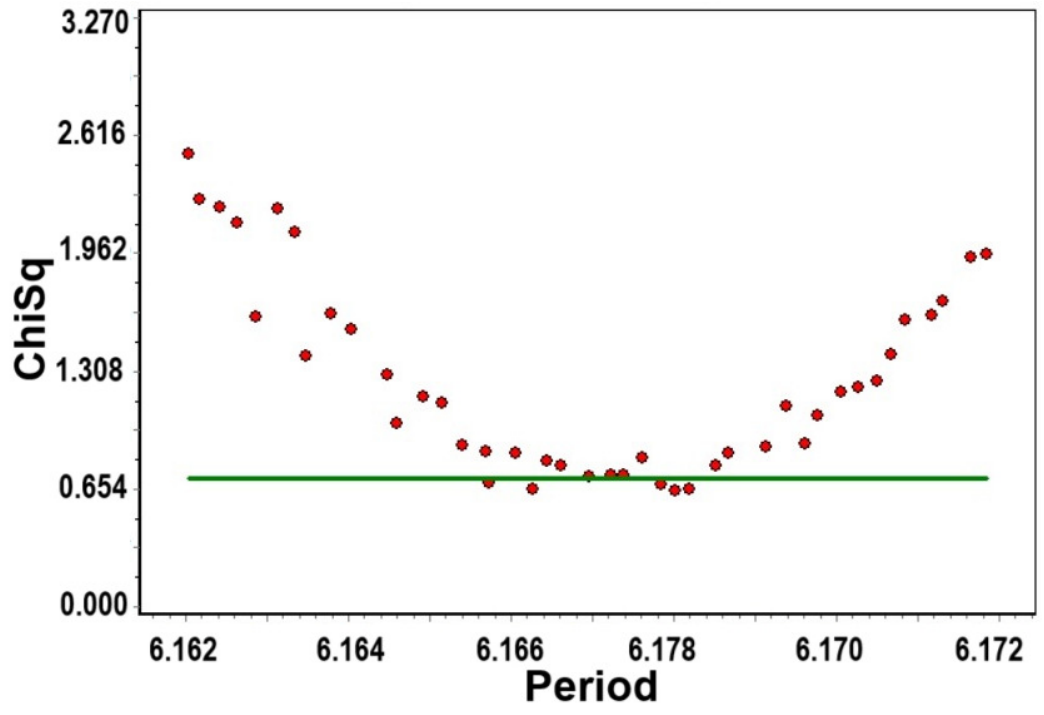}
\caption{The periodograms of the initial sidereal period search around our synodic period of 6.172\,h.}\label{fig:modelPRDGM}
\end{figure}

Even though the irregular form of the observed lightcurve we decided to test for a single-body low-resolution shape model. For this purpose, we used the lightcurve inversion method \citep{Kaas1,Kaas2} and LCInvert programme part of the MPO Software\footnote{\url{https://minplanobs.org/BdwPub/php/displayhome.php}}. Considering the tremendous accuracy of the DR3 photometric data we took the weighting factor to be set to 1.0 for both sparse and for dense photometric data. 
These sparse data added to our dense data give us a wide range of the PAB longitude distribution (see Fig.~\ref{fig:aspect}) which is very important for the shape reconstruction.
We made the initial sidereal period search around our synodic period of 6.172\,h (see Fig.~\ref{fig:modelPRDGM}--left panel). A narrower period search found a few solutions with 10\,\% of the lowest $\chi^2$ (see Fig.~\ref{fig:modelPRDGM}--right panel), but we choose the one with the lowest $\chi^2$, which is 6.16802461\,h, as the most likely period and used it for the pole search. 

\begin{figure*}[h]%
\centering
\includegraphics[width=0.8\textwidth]{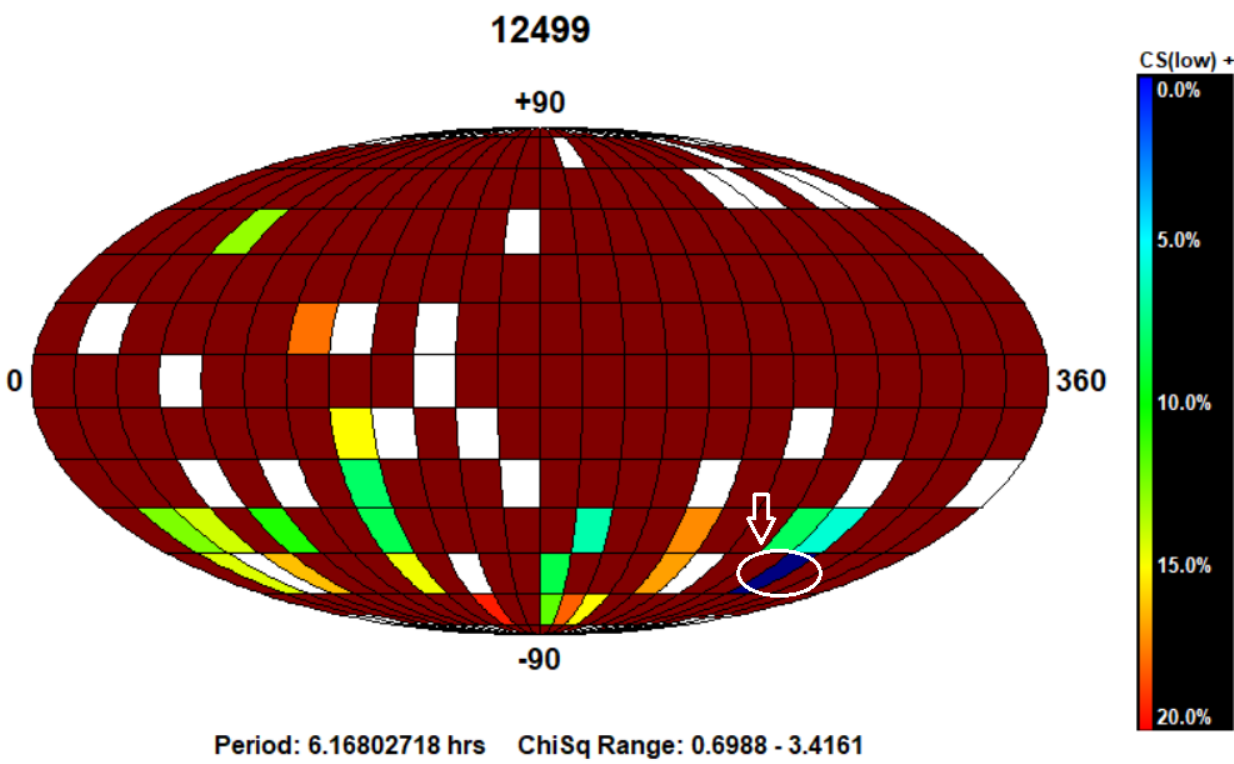}
\caption{The $\chi^2$-map of the pole search distribution. The white arrows indicate the dark blue region with the smallest value of $\chi^2$. The dark blue regions have $\chi^2$ values about 20\% smaller than $\chi^2$ values of the dark red regions.}\label{fig:ChiMap}
\end{figure*}

\begin{figure*}[h]%
\centering
\includegraphics[width=0.7\textwidth]{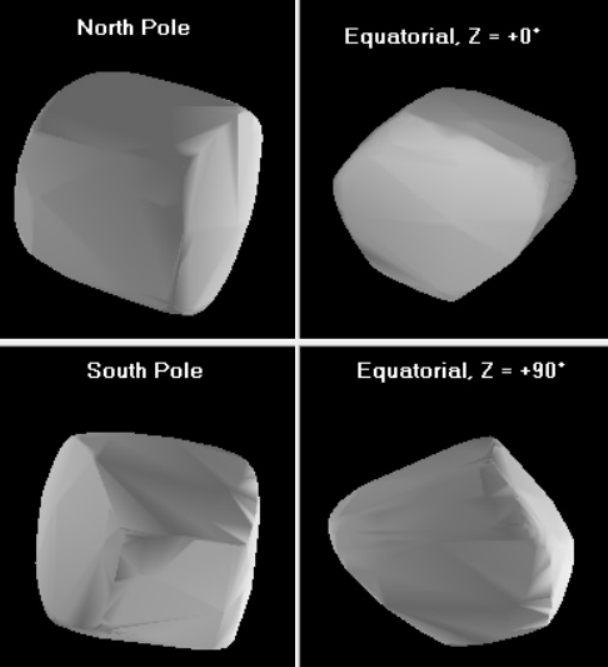}
\caption{The 3D low-resolution convex shape model of (12499) 1998 FR47. The north and south pole views are given in two left panels, while the equatorial viewings with rotational phases 90$^\circ$ apart are given in two right panels. The calculated aspect ratio of the longest to two shorter semi-axis is a/b=1.1 and a/c=1.3.}\label{fig:Shape}
\end{figure*}

\begin{figure*}[h]%
\centering
\includegraphics[width=0.8\textwidth]{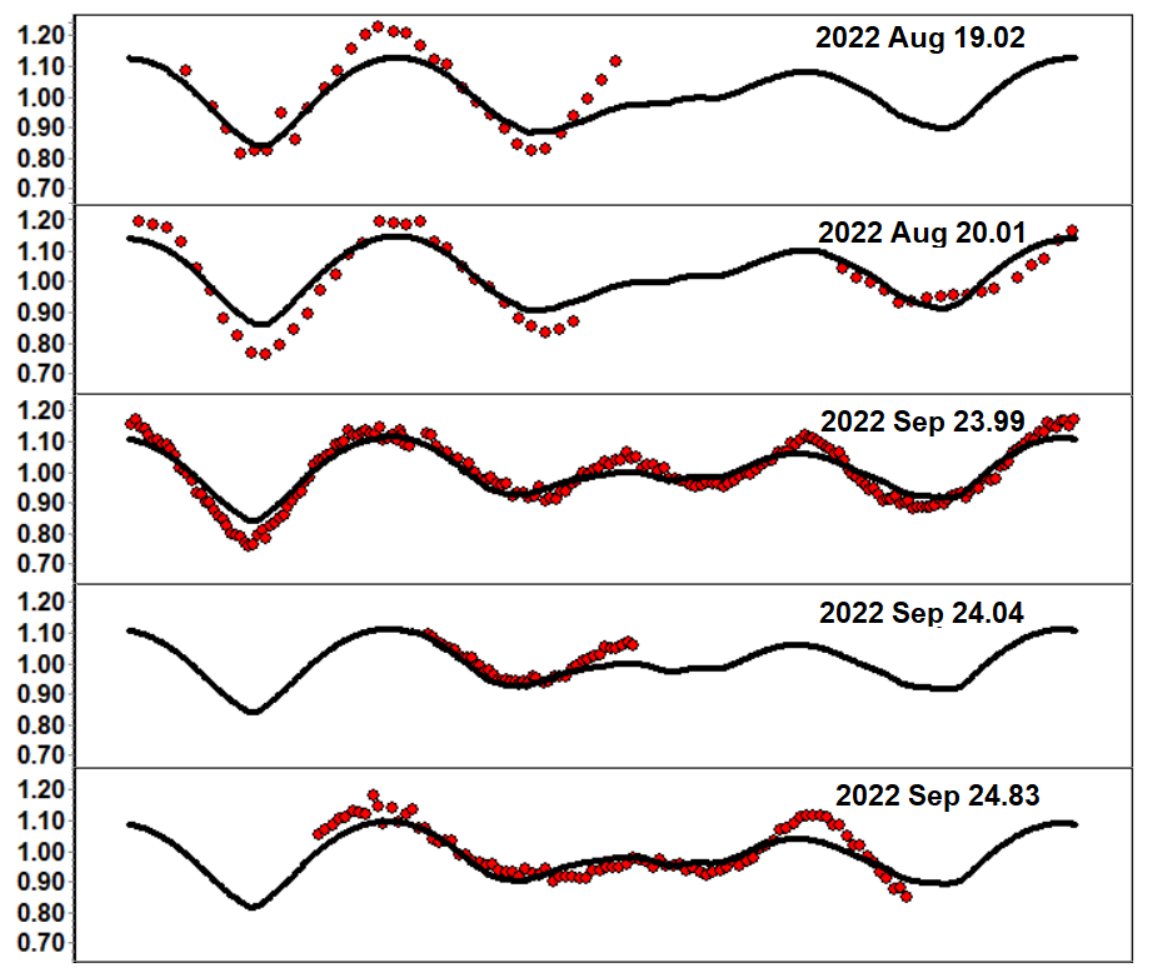}
\caption{The lightcurve (red points) obtained from our dense observations superimposed over the lightcurves generated by the model constructed using a combination of our dense and GAIA DR3 sparse data.}\label{fig:modelLC}
\end{figure*}

The pole search produced 312 discrete and fixed pole positions distributed over the unit sphere with 15$^\circ$ steps in ecliptic latitude and longitude. The dark blue region in the constructed $\chi^2$-map, presented in Fig.~\ref{fig:ChiMap}, has $\chi^2$-value about 25\,\% smaller than $\chi^2$-values of the dark red region. It corresponds to pole coordinates $\lambda$=300$^\circ$ and $\beta$=-60$^\circ$ and sidereal period 6.168027718\,h. Taking those values as free parameters we made a finer pole search and we got sidereal period 6.16799606\,h and pole coordinates $\lambda$=291.9$^\circ$ and $\beta$=-53.2$^\circ$, which suggest the retrograde sense of the asteroid rotation. Assuming the triaxial ellipsoid form of the asteroid with axes a$>$b$>$c, which rotates around the shortest axis c=1, the calculated pole coordinates corresponded to the relative shape dimensions a/b=1.1 and a/c=1.3. The corresponding 3D shape model is also constructed and presented in Fig.~\ref{fig:Shape}. The comparison of the lightcurves generated from this model with the dense lightcurves is shown in Fig.\ref{fig:modelLC}. The modelled lightcurves in Fig.\ref{fig:modelLC} do not correspond well enough with the observations. 
These, together with the initial irregular form of the observed lightcurve, were indicators for searching for a double-period solution caused by the binary nature of the asteroid.

\subsection{Possible binary system}\label{sec:bin}

\begin{figure*}[h]%
\centering
\includegraphics[width=0.6\textwidth]{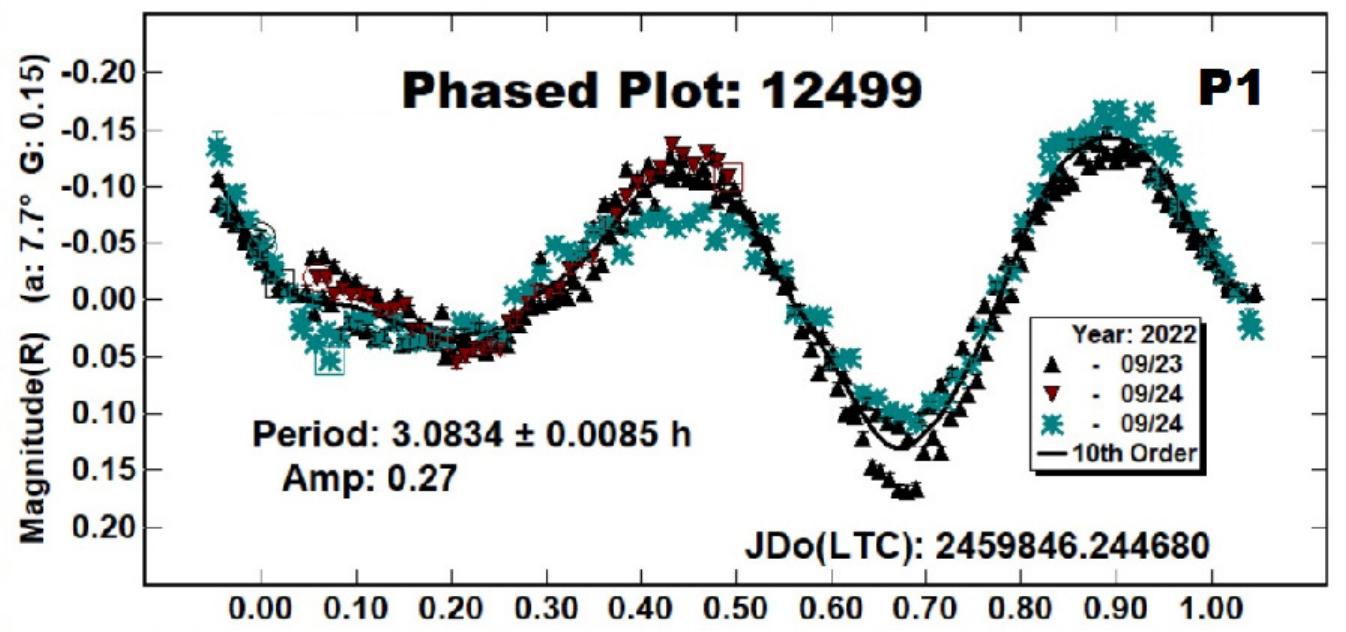}
\includegraphics[width=0.39\textwidth]{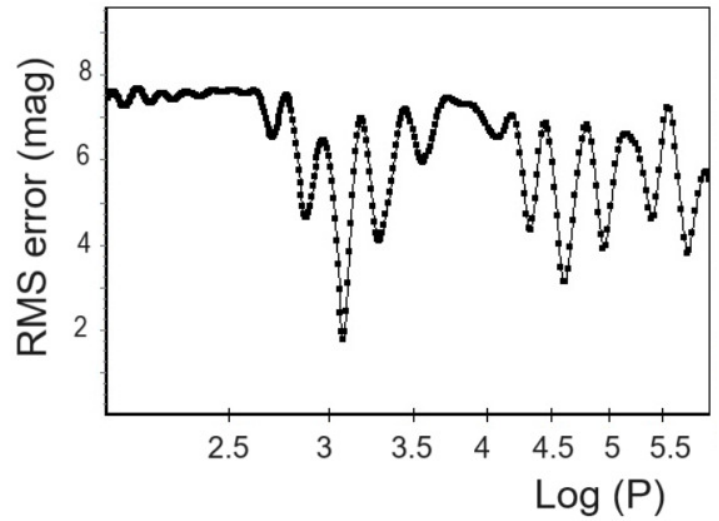}
\caption{(12499) 1998 FR47 -- the lightcurve of the primary (left panel) and the corresponding dual-period search periodogram (right panel).}\label{fig:primary}
\end{figure*}

\begin{figure*}[h]%
\centering
\includegraphics[width=0.6\textwidth]{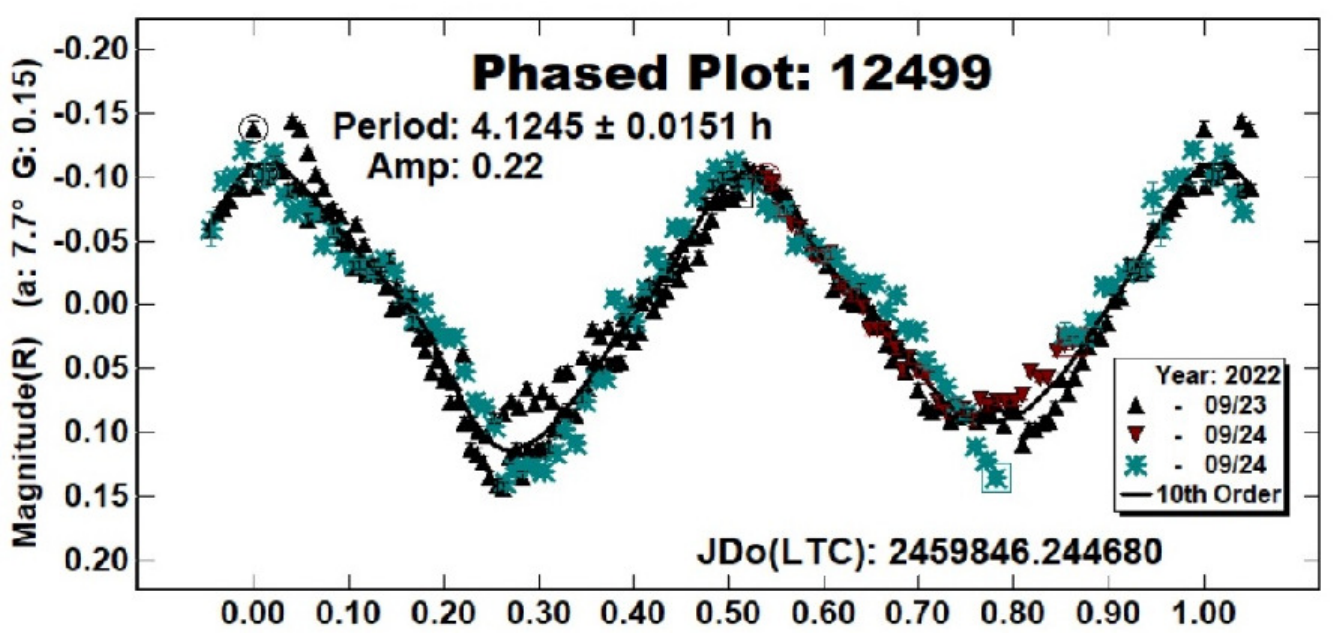}
\includegraphics[width=0.39\textwidth]{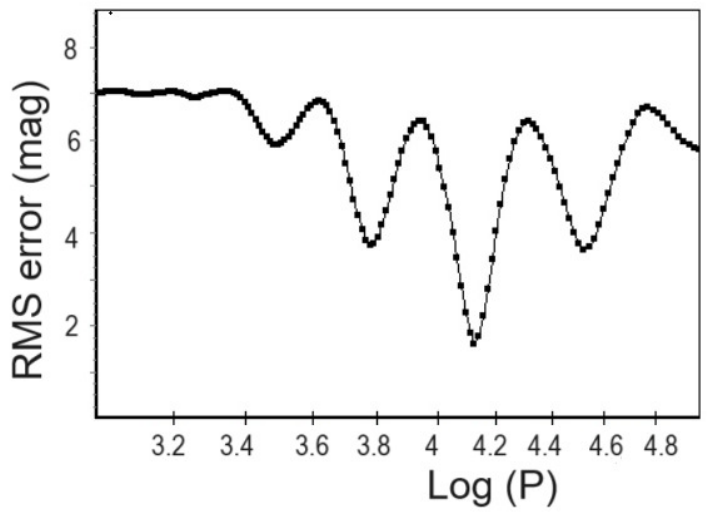}
\caption{(12499) 1998 FR47 -- the lightcurve of the secondary (left panel) and the corresponding dual-period search periodogram (right panel).}\label{fig:secondary}
\end{figure*}

The complex form of the lightcurves, the unusual depth of one of the minima, as well as the diameter of the asteroid (smaller than 10\,km), suggest that there could be two additive components in the lightcurve and that (12499) could be a candidate for a binary asteroid \citep{Pravec2019}. Using a Dual-period search \citep{Warner.book} in MPO Canopus, we calculated the rotational period of the synodic period of both components. The first one has a period of 3.0834$\pm$0.0085\,h and an amplitude of 0.27\,mag and is presented in Fig.~\ref{fig:primary}. The second one is with a period of 4.1245$\pm$0.0151\,h and an amplitude of 0.22\,mag and is presented in Fig.~\ref{fig:secondary}. 
Unfortunately, our observations did not capture any superimposed mutual events, and the distance between the two bodies is not constrained. The orbital period of the satellite using the data from these events could not be determined, which suggests that it may be a wide binary system.

Additional data from our observations and data from observers sufficiently far away in longitude are needed to cover the orbital period with mutual events (i.e. occultations and/or eclipses). It will help not only to decide if (12499) is a binary asteroid but also to refine our calculated periods for the primary and for the secondary, as well as to determine the orbital period of the system. The best way is to organise an observational campaign on this object in the future.

\subsection{Taxonomy from Gaia spectroscopy}\label{sec:spec}

The Gaia mission has been observing Solar System objects since the beginning of its operation\citep{GaiaSS}. The Gaia DR3 includes, for the first time, the mean reflectance spectra of selected asteroids\citep{GaiaSSsp}. 
Considering that (12499) is among them, we will use that data to further investigate its properties in the direction of taxonomy and composition, relying on the new probabilistic approach by \citet{classy}. 

\begin{figure*}[h]%
\centering
\includegraphics[width=0.9\textwidth]{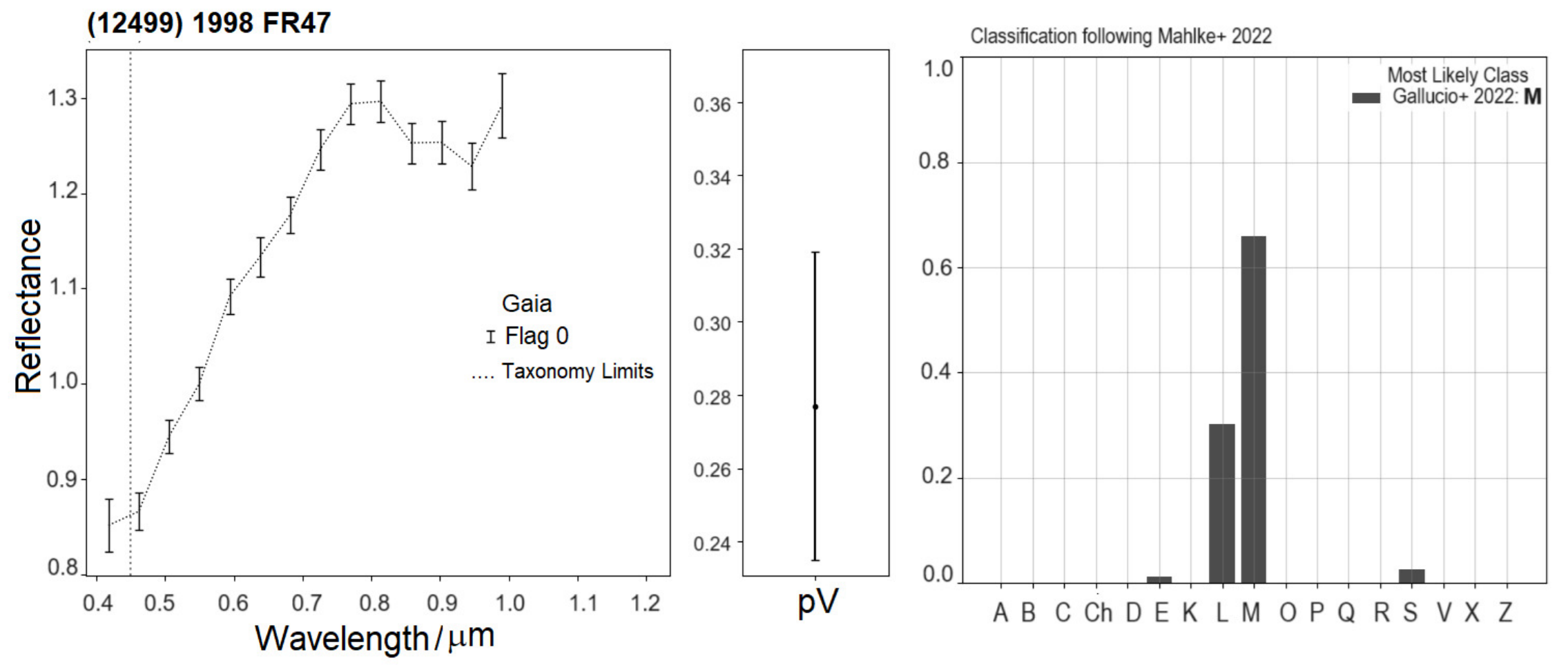}
\caption{Gaia DR3 reflectance spectrum of (12499) 1998 FR47 and its taxonomical classification according to \citet{classy} (see text for details).}\label{DR3_sp}
\end{figure*}

The results are shown in Fig.~\ref{DR3_sp}. The left panel shows the Gaia DR3 reflectance spectra of the object, the middle panel points out its latest published albedo obtained by NEOWISE \citep{NEOWISE}, and the right panel presents the possible taxonomy classes with their probabilities. 

In particular, the asteroid has a 66\,\% probability of being an M-class object and 30\,\% for an L-class. Both of these solutions fall into very diverse classes situated between primitive carbonaceous C- and silicate (partially thermally metamorphosed) S-classes \citep{Cloutis90,Cloutis11,Vernazza14}. Other low probability solutions are 3\,\% for being an S-type object and 1\,\% for E-type.

The most probable  M-class, which according to \citet{classy} corresponds to the old \citet{Tholen84} M-class or the newest Bus-DeMeo \citep{Bus02,DeMeo09} Xk-class. Its reflectance spectrum has a linear red slope and possible faint features around 0.9\,$\mu$m. The objects of this class are associated with at least two populations, the chondritic and the metallic \citep{Vernazza11,Viikinkoski17}, whose composition can vary, depending on their distance from the Sun and the time of format. The most probable formation scenario for M-type objects is that they are formed from a differentiated parent body after a collision and the fragments from the core, which were rich in metals, were formed them. 

On the other hand, the reflectance spectrum of L-class asteroids is variable apart from a red visible slope and a small feature around 1\,$\mu$m \citep{Sunshine08}. The composition of this type of object is associated with large abundances of spinelbearing calcium–aluminium-rich inclusions, suggesting they may be pieces of the first planetesimals to form in the Solar System's protoplanetary disk \citep{Ltype}.

\subsection{Dynamical properties}\label{sec:dyn}

 Let us first remark that (12499)'s membership to the Flora family indirectly suggests that (12499) could be stable in long times, since the age of the family is estimated to $\sim 1$\,Gyr \citep{2013A&A...559A.134H, 2014Icar..243..111D, 2013A&A...551A.117B}. We will verify these stability assumptions in two ways: (a) by mapping and studying the orbital neighbourhood of the asteroid (b) by tracking the evolution of (12499) and several fictive objects in its close neighbourhood. 

The calculation of maps is performed using the Fast Lyapunov Indicator - FLI \citep{1997CeMDA..67...41F}, a numerical tool which is efficient in dynamical cartography of both realistic \citep{2016CeMDA.124..335D, 2022CeMDA.134....6D, 2017MNRAS.464.4063R, 2020IAUGA..30...17T, 2017MNRAS.465.4441T} and idealised systems \citep{2000Sci...289.2108F}.  FLI values and the orbital evolution of individual test particles (TPs) are obtained with the orbit9 integrator available at \url{http://adams.dm.unipi.it/orbfit/}. We consider the gravitational influences of seven planets from Venus to Neptune. The mass of Mercury is added to the mass of the Sun, and a proper barycentric correction is applied. The non-gravitational Yarkovsky and YORP effects are not included in the calculation. In the integration, asteroids are treated as massless test particles. We do not consider their sense of rotation, nor the eventual mutual gravitational effect of the components in the possible binary system.

\begin{figure}[h]
\includegraphics[width=\textwidth]{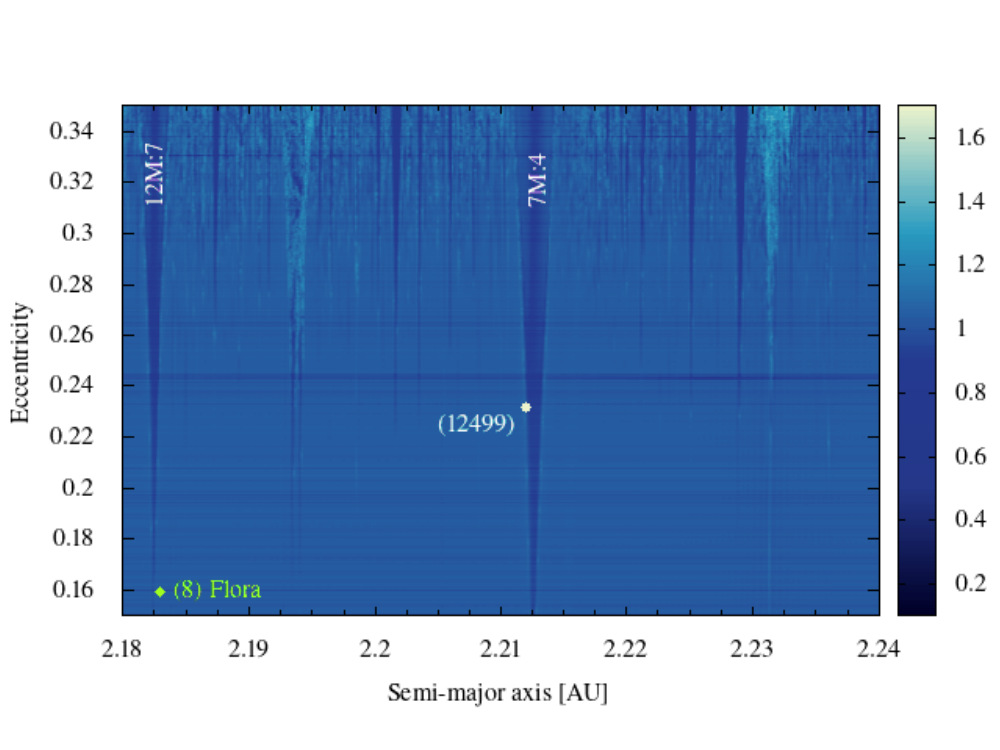}
\caption{The 5000 years FLI map calculated in the semi-major axis, eccentricity plane for $[a,e]=[2.18,2.24]\times[0.15,0.35]$, where most of the vertical V-shaped structures are mean motion resonances with Mars. The asteroid (12499), marked with a white dot, lies close to the 7M:4 MMR  at $a \sim 2.212$\,AU. The position of the asteroid (8) Flora, the largest body in the family, is marked with a green diamond-like shape close to the 12M:7 MMR at about $a \sim 2.182$ AU.} 
\label{fig11}
\end{figure}

Fig.~\ref{fig11} shows one segment of the semi-major axis eccentricity  $(a, e)$ plane around the asteroid (12499). It is produced by taking 250\,000 TPs regularly distributed on a 500\,x\,500 grid in $[a,e]=[2.18,2.24]\times[0.15,0.35]$. For each TP we calculate its corresponding FLI for 5000 years. Stable orbits have lower FLIs and are marked with dark blue colours, while the more chaotic ones have higher FLIs and are coloured with lighter shades of blue. The four remaining orbital elements, the inclination $i$,  the longitude of the node $\Omega$,  the argument of the pericentre $\omega$, and the mean anomaly $M$, are equal to the corresponding elements of (12499) for the epoch 2458800.5 MJD (13. November 2019). Their values are taken from the JPL NASA database\footnote{Available at \url{https://ssd.jpl.nasa.gov/}} and are approximately  $(i, \Omega, \omega, M)\sim (6.76^{\circ}, 337.11^{\circ}, 6.71^{\circ}, 53.79^{\circ})$. By using this mapping method, we can see the realistic dynamical environment of (12499).

The map in Fig.~\ref{fig11} shows a large number of mean motion resonances (MMRs) appearing as vertical thin V-shaped structures. Additional calculations (not shown here) verified that most of them are resonances with Mars, confirming the domination of such resonances in the Flora family  \citep{2017AJ....153..172V, 2002Icar..157..155N}.  

The location of the asteroid (12499) at $(a,e)$ $=(2.21209$, $0.23150)$ is marked with a white dot, while the green diamond shape at $(a,e)=(2.18289, 0.15945)$ shows the position of the asteroid (8) Flora, the possible largest remnant of the parent body of the family. Although our paper focuses on the asteroid (12499), we will make some other remarks on Fig.~\ref{fig11}. \cite{2017AJ....153..172V} showed that the asteroid (8) Flora is close to the $\nu_6$ and $\nu_{16}$ secular resonances\footnote{Resonances $\nu_6$ and $\nu_{16}$ occur when the precession and nodal frequencies of the asteroid's orbit equal the precession and nodal frequencies of the orbit of Saturn, respectively.}. Their presence is not visible on our map since it would require much longer integration times than 5000 yrs used here (see \cite{2015MNRAS.451.1637T} for the detection of secular resonance in FLI maps).  In Fig.~\ref{fig11} it appears that (8) Flora is close to the 12:7 MMR with Mars at about $a\sim 2.182$ AU (for the identification of mean motion resonances see e.g. \citet{2006Icar..184...29G, 2023A&C....4300707S}). Still, its exact distance from the 12M:7 resonance as visible on the chart, is not entirely reliable, since the map illustrates the dynamical situation in the orbital plane of the asteroid (12499). The shape and the positions of resonances change with the change of orbital angles, and according to the JPL database for the particular epoch, (8) Flora is 199.6 degrees in front of (12499) in the mean anomaly $M$, it has 226.2 degrees larger longitude of the node $\Omega$, and is 79.6 degrees behind in the argument of pericentre $\omega$, id est $M_{(8)} - M_{(12499)} = 199.6^{\circ},  \Omega_{(8)} - \Omega_{(12499)}= 226.2^{\circ}, \omega_{(8)} - \omega_{(12499)}= - 79.6^{\circ}$. 

\begin{figure}[h]
\centering
\includegraphics[width=0.9\textwidth]{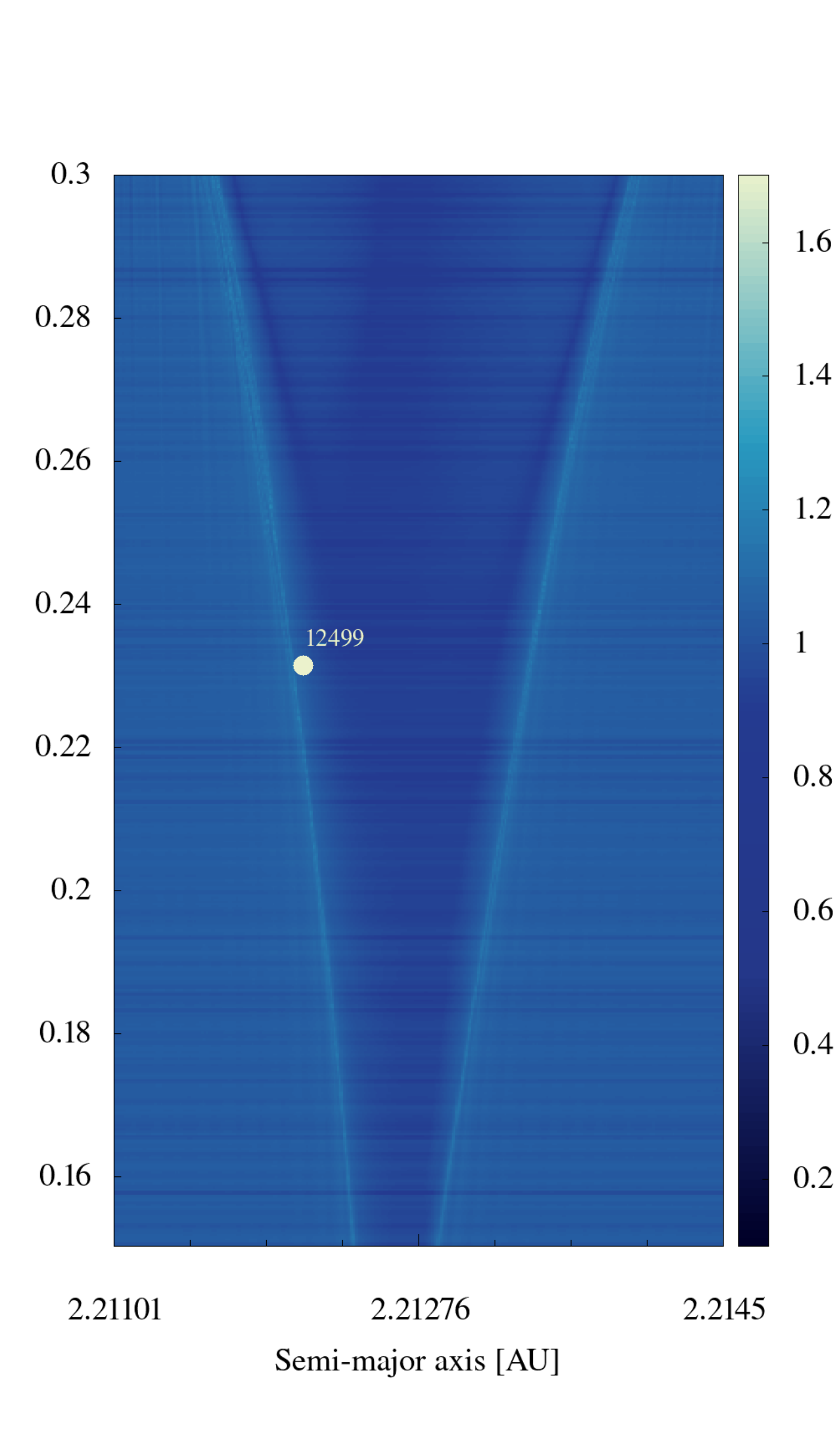}
\caption{The enlarged portion of Fig.~\ref{fig11} around the 7M:4 resonance for $[a,e]=[2.21101, 2.2145]\times[0.15,0.3]$. The asteroid (12499), marked with the white dot, is located exactly on the chaotic border of the resonance.} 
\label{fig12}
\end{figure}

In Fig.~\ref{fig11} the asteroid (12499) is placed at the 7:4 mean motion resonance with Mars. In a closer look given in Fig.~\ref{fig12}, for $[a,e]=[2.21101, 2.2145]\times[0.15,0.3]$, we notice that (12499) lies exactly on the chaotic border of the resonance. 

Several asteroid pairs are already associated with mean motion resonances (see for example \cite{Pravec2019}, \cite{Duddy2012}, or \cite{2024arXiv240303559R}). We are unaware that any such pair is found on the very edge of resonance, as is the case for (12499). It remains an open question whether the position on the resonant border may be related to the duality (binarity) of the asteroid.

\begin{figure}[h]%
\centering
\includegraphics[width=\textwidth]{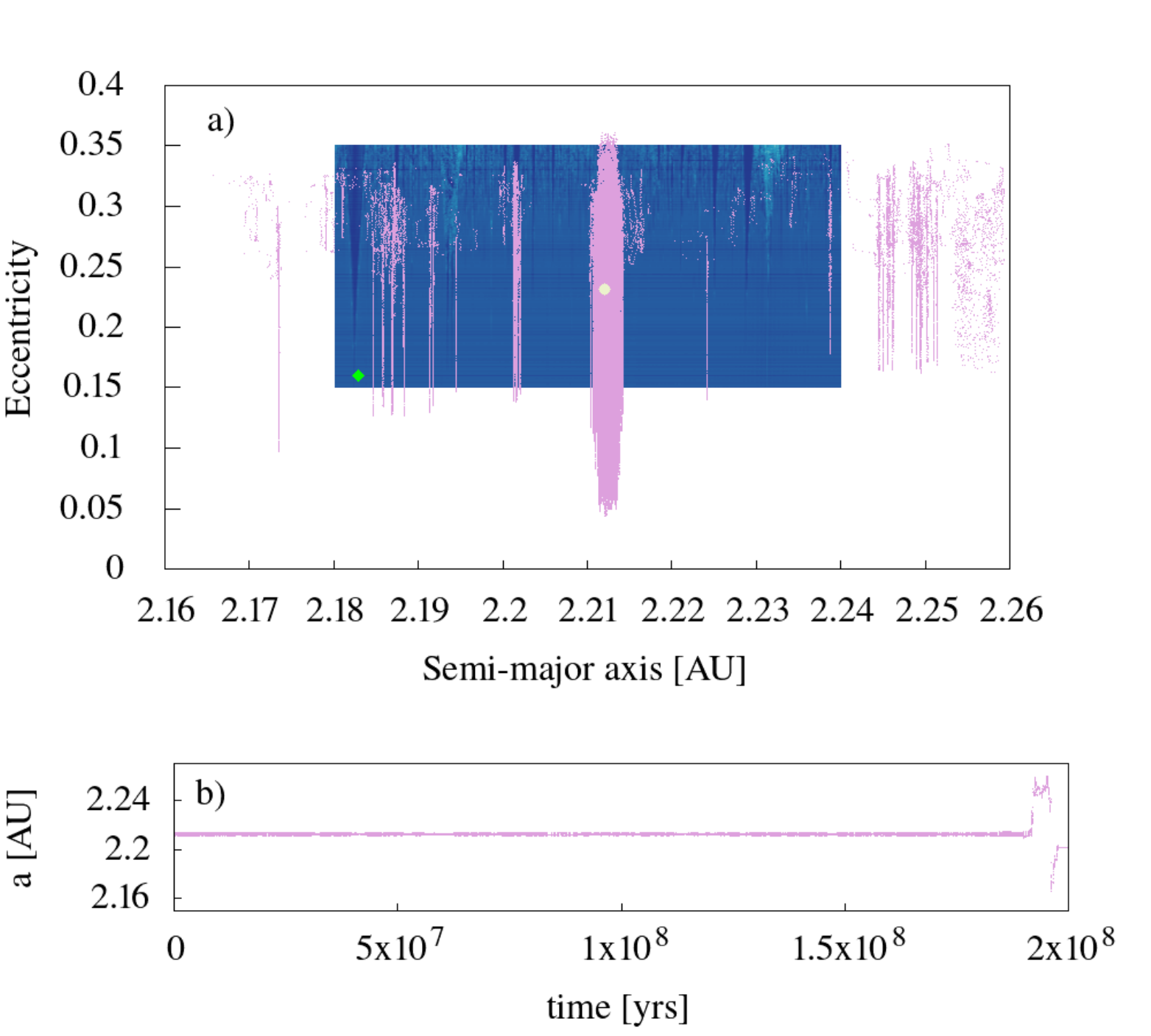}
\caption{The top panel shows the orbital evolution of (12499) in the $(a,e)$ plane during 200\,Myrs of integration. The places visited by the asteroid are coloured pink. The map from Fig.~\ref{fig11} is set in the background for better orientation. In addition to the 7M:4 resonance at $a\sim 2.212$\,AU (from which it originated), (12499) visited numerous neighbouring resonances but did not leave the Flora family. The lower panel shows the evolution of $a_{(12499)}$ during the whole integration time. The asteroid remains close to (or inside) the 7M:4 resonance at $a \sim 2.212$\,AU for almost 190\,Myrs. Its excursion to the neighbouring MMRs occurred in the last 10\,Myrs of integration.} 
\label{fig13}
\end{figure}

In the following part, we will verify the long-term stability of (12499). In Fig.~\ref{fig13} we show two perspectives of its orbital evolution during 200\,Myrs. The top panel shows which parts of the $(a,e)$ plane are visited by (12499), where its trace is coloured pink. The lower panel gives the change of (12499)'s semi-major axis $a_{(12499)}$ during the integration time. The value of $a_{(12499)}$ remains close to the centre of the 7M:4 resonance at $\sim 2.212$\,AU for about 190\,Myrs, after which it starts an excursion to nearby resonances, but inside the family. We use the rough boundaries of the Flora family from \cite{2002Icar..157..155N} which are $2.12 < a < 2.31$\,AU, $0.11 < e < 0.175$, and $3 < i <7.5$. The (12499)'s perihelion distance $q$ reached Mars crossing values  ($q=1.665$\,AU) several times, for the first time after about 0.6\,Myrs. However, (12499) did not become a Near Earth Object ($q<1.3$\,AU) since its minimal perihelion distance was about $q \sim 1.4$\,AU.

\begin{figure}[h]%
\centering
\includegraphics[width=\textwidth]{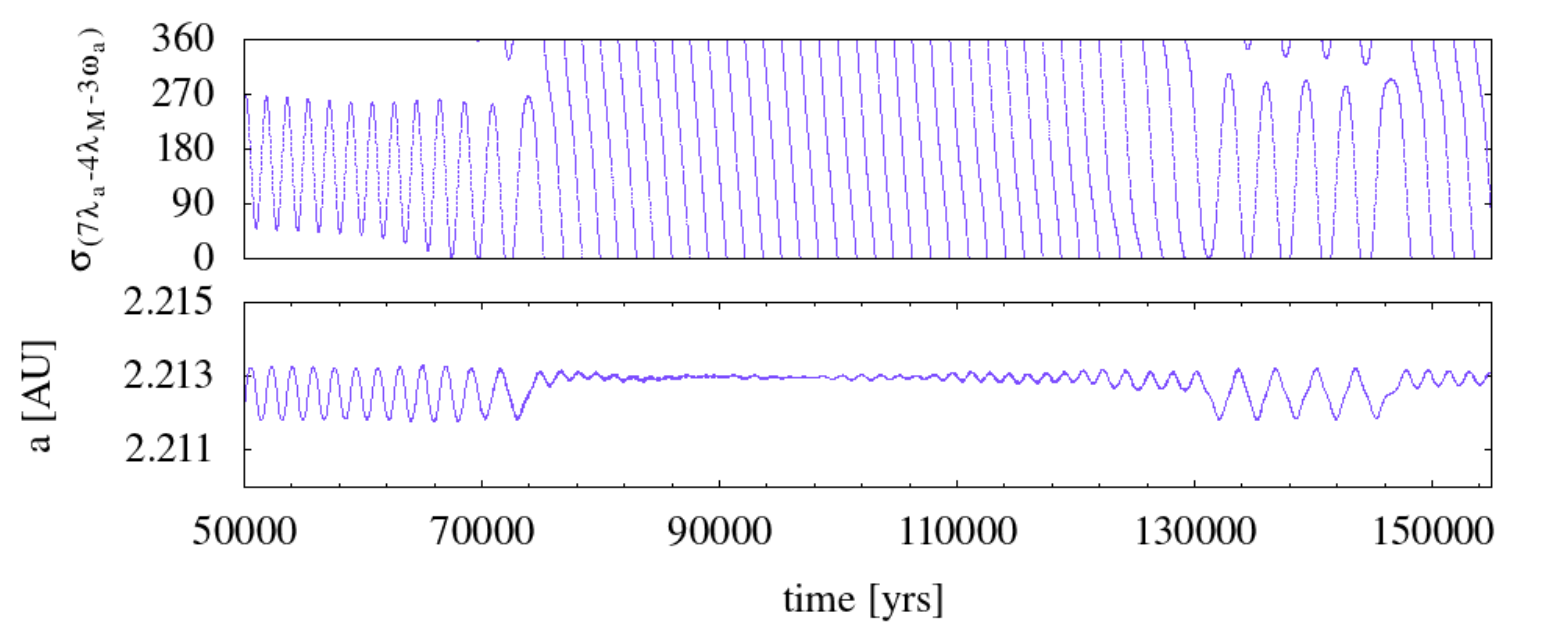}
\caption{The change of (12499)'s critical angle for the 7M:4 MMR $\sigma_{7M:4}$ (top panel) and the change of its semi-major axis $a_{(12499)}$ (bottom panel) for $t\in [50, 155]$\,Kyrs. The asteroid spent around 72\,Kyrs librating in the resonance (the first 50\,Kyrs are not presented). In the next 80\,Kyrs i.e. until $\sim 130$\,Kyrs, $\sigma_{7M:4}$ circulates, meaning the asteroid was pushed out of the resonance, but not far since its semi-major axis has a quasi-constant value close to the resonance at about $a_{(12499)} \sim 2.213 \,AU$.  Another episode of libration in $\sigma_{7M:4}$ and oscillation in $a_{(12499)}$ occurred between {130 and 150\,Kyrs}, meaning the asteroid is trapped in the resonance again.  We observed that such libration-oscillation modes in $\sigma_{7M:4}$ alternate for about 190 million years.}
\label{fig14}
\end{figure}

Another look at (12499)'s evolution  for $t \in [50, 155]$\,Kyrs is given in Fig.~\ref{fig14}. On the top plot, we track the critical angle of (12499) for the 7M:4 resonance defined with $\sigma_{7M:4}=7\lambda_a + 4\lambda_M - 3{\Tilde{\omega}_a}$, where $\lambda_a$ and $\lambda_M$ are the mean longitudes of the asteroid and Mars respectively, (defined with $\lambda = M + \omega + \Omega$), while $\Tilde{\omega}_a$ is the asteroids longitude of perihelion $\Tilde{\omega} = \omega + \Omega$. The lower panel shows the evolution of $a_{(12499)}$ for the same time. We notice that $\sigma_{7M:4}$ and $a_{(12499)}$ have synchronised changes. In the first 80\,Kyrs, $\sigma_{7M:4}$ librates, and $a_{(12499)}$ oscillates, meaning the asteroid is trapped in the resonance. In the next 70\,Kyrs the asteroid is outside the resonance since the critical angle circulates (it has smooth changes in the $[0,360]$ interval), while $a_{(12499)}$ is almost constant. The next entry in the 7M:4 resonance starts at $\sim 130$\,Kyrs and lasts for about 20\,Kyrs. Such alternating episodes of going in and out of the resonance are observed for 190\,Myrs. In Fig.~\ref{fig14} we plotted a relatively short time interval of about $\sim 100$\,Kyrs (between $t \in [50, 155]$\,Kyrs), because the fine oscillations in $\sigma_{7M:4}$ and $a_{(12499)}$ are not noticeable for a 200\,Myrs interval (see panel b in Fig.~\ref{fig13}). 

In the end, we selected 20 additional TPs from the chaotic border of the 7M:4 MMR from Fig.~\ref{fig12} and observed their orbital evolution in 50\,Myrs. Assuming (12499) could be a binary asteroid lying on the chaotic border of a resonance, we may expect that initially close segments could have a large orbital divergence. We emphasise again that we did not treat the gravitational interaction between the possible components of the binary, but only the evolution of individual massless TPs in the above-described Solar system model.

\begin{figure}[h]%
\centering
\includegraphics[width=\textwidth]{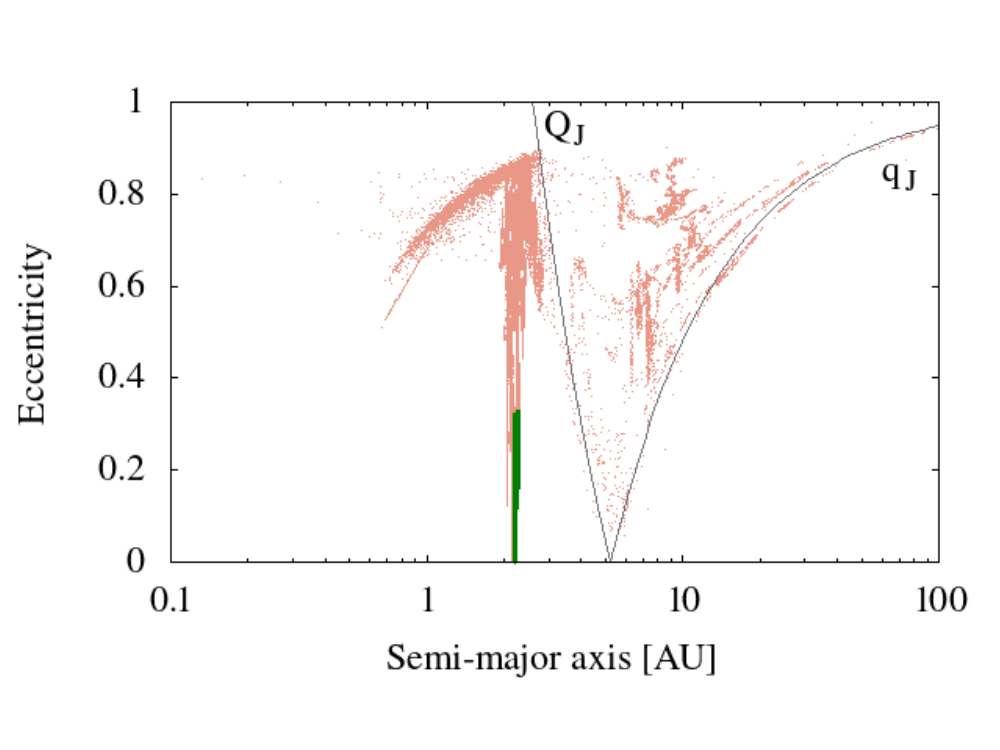}
\caption{The trace of 20 test asteroids initially placed on the borders of the 7:4 MMR with Mars, during 50\,Myrs of orbital evolution. We cover a large segment of the Solar system  $[a,e]=[0,100]\times[0,1]$, where the semi-major axis is given in the $\log_{10}$ scale. Among the 20 test objects, four of them were thrown out of resonance due to close approaches with Mars (and other inner planets), entered the NEO region and experienced a macroscopic diffusion through the Solar system (their traces are coloured red). The 16 remaining bodies stayed close to the 7M:4 MMR around $a\sim 2.212 $\,AU. Their trace in the chart is coloured green. The two apsidal lines of Jupiter $q_J$ and $Q_J$ are marked with thin light grey lines.}
\label{fig15}
\end{figure}

The results of this last set of integration are given in Fig.~\ref{fig15}. Again, we show the $(a,e)$ plane but for a larger domain for $[a,e]=[0, 100] {\rm AU} \times [0,1]$, where the semi-major axis is presented on the logarithmic scale. This portion of the $(a,e)$ plane was selected because we observed large-scale changes in the orbital elements of the test asteroids.
More precisely, 4 among the 20 objects experienced macroscopic evolution through the Solar system. These four asteroids were thrown out of the resonance by close approaches with Mars, followed by a series of encounters with Earth and Venus. Only one among the four test objects had an approach with Jupiter. The resulting orbital pathways in the $(a,e)$ plane are coloured in red (Fig.~\ref{fig15}).
We notice they spread mostly between the apsidal lines of Jupiter $q_J$ and $Q_J$ (marked with two grey lines). The first entries in the NEO region are observed at about 20\,Myrs, in significantly shorter times than the average deliveries of 100\,Myrs as reported in \cite{2002Icar..157..155N}. The remaining 16 test asteroids stayed in the 7M:4 resonance; their trace in the chart at about 2.21 AU, is coloured green. Let us mention that this result represents a possible outcome of the chaotic evolution, whereas, in reality, we could expect shorter evolution timescales, due to eventual close approaches with the largest asteroids and/or nongravitational effects like YORP and Yarkovsky.

Our results show that the Flora family is efficient in supplying the NEO region also through weak MMRs with Mars, not only via the $\nu_6$ resonance as a main driver. Still, we emphasise that this last set of test asteroids is intentionally chosen from the resonant chaotic border, meaning the real possibilities for this scenario are somewhat lower, but not impossible. 

\section{Conclusions}\label{sec:con}

The unusual shape of the observed lightcurve for the asteroid (12499) 1998 FR47 was investigated in different senses. Firstly we try to determine the rotational period, axis poles, and shape of the object assuming a single body. The solution suggests a rare quadramodal shape of the lightcurve with a rotational period of 6.172$\pm$0.003\,h and an unusual shape to model it. Then we saw that the lightcurve can be modelled by two different periods and performed a dual-period search assuming that the asteroid is a binary system. The two solutions are 3.0834$\pm$0.0085\,h and 4.1245$\pm$0.0151\,h respectively for the primary and secondary bodies. No mutual event was detected in the lightcurve, so we suppose that these periods are of the individual lightcurves of the two bodies of the system, which is not synchronised and most probably is a wide binary system.

The available Gaia DR3 reflectance spectrum of the object suggests either M- or L-type taxonomy. They both are from the M-complex and suggest either chondritic or metallic composition or an object with large abundances of spinelbearing calcium–aluminium-rich inclusions, suggesting that the asteroid (12499) might be a piece of the first planetesimals to form in the Solar System protoplanetary disk.

Dynamical studies showed that (12499) is located at the very edge of the 7:4 mean motion resonance with Mars, where it should spend the next 190\,Myrs, after which it should start a tour of nearby resonances. Additional calculations showed that TPs on the chaotic border of the 7M:4 MMR, close to (12499), may experience shorter stays in the resonance and faster deliveries to the NEO region and/or macroscopic diffusion starting already at 20\,Myrs. Since our model does not include Yarkovsky or YORP effects, there is a possibility that the reported times are even shorter. 

\section{Acknowledgements}
The authors gratefully acknowledge observing grant support from the Institute of Astronomy and Rozhen National Astronomical Observatory, Bulgarian Academy of Science. The observations at AS Vidojevica and funding for N.T. were financed from the Ministry of Education, Science, and Technological Development of the Republic of Serbia (contract no. 451-03-66/2024-03/200002). The authors are thankful to Petr Pravec for the helpful discussion of the possible binary origin of the asteroid, and to the anonymous reviewer for the useful insights and suggestions.
This work has made use of data from the European Space Agency (ESA) mission
{\it Gaia} (\url{https://www.cosmos.esa.int/gaia}), processed by the {\it Gaia}
Data Processing and Analysis Consortium (DPAC,
\url{https://www.cosmos.esa.int/web/gaia/dpac/consortium}). Funding for the DPAC
has been provided by national institutions, in particular the institutions
participating in the {\it Gaia} Multilateral Agreement.

\bibliography{12499-bibliography}
\end{document}